\documentclass[aps,prd,onecolumn,nofootinbib,groupedaddress,superscriptaddress]{revtex4}
\usepackage{graphicx}
\usepackage{epstopdf}
\usepackage{amsmath}
\usepackage{amsfonts}
\usepackage{amssymb}
\usepackage{appendix}
\usepackage{enumerate}
\usepackage{comment}
\usepackage{bbold}
\usepackage[shortlabels]{enumitem}
\usepackage{color}
\usepackage{slashed}
\usepackage{subfigure}
\usepackage{setspace}
\usepackage{footnote}
\usepackage{multirow}
\usepackage{longtable}
\usepackage[colorlinks = true,
            linkcolor = blue,
            urlcolor  = blue,
            citecolor = blue,
            anchorcolor = blue]{hyperref}
\usepackage[capitalize]{cleveref}
\usepackage{braket}
\usepackage{multirow}
\usepackage{mathtools}
\usepackage{physics}

\usepackage{etoolbox} 
\AtBeginEnvironment{pmatrix}{\everymath{\displaystyle}}

\usepackage{xspace}

\makeatletter
\DeclareRobustCommand\onedot{\futurelet\@let@token\@onedot}
\def\@onedot{\ifx\@let@token.\else.\null\fi\xspace}

\def\eg{\emph{e.g}\onedot} 
\def\ie{\emph{i.e}\onedot} 
\makeatother

\usepackage[a4paper, portrait, margin=1in]{geometry}

\begin{document}

\singlespacing

{\hfill CERN-TH-2022-116}

{\hfill FERMILAB-PUB-22-508-T}

{\hfill MS-TP-22-19}

{\hfill NUHEP-TH/22-07}

\title{Gravitational Waves from Current-Carrying Cosmic Strings}

\author{Pierre Auclair}
\email{pierre.auclair@uclouvain.be}
\affiliation{Institute of Mathematics and Physics, Louvain University, 1348 Louvain-la-Neuve, Belgium}

\author{Simone Blasi}
\email{simone.blasi@vub.be}
\affiliation{Theoretische Natuurkunde and IIHE/ELEM, Vrije Universiteit Brussel, \& The  International Solvay Institutes, Pleinlaan 2, B-1050 Brussels, Belgium}

\author{Vedran Brdar}
\email{vbrdar@fnal.gov}
\affiliation{Theoretical Physics Department, Fermilab, P.O.\ Box 500, Batavia, IL 60510, USA}
\affiliation{Northwestern University, Department of Physics and Astronomy, Evanston, IL 60208, USA}

\author{Kai Schmitz}
\email{kai.schmitz@uni-muenster.de}
\affiliation{University of M\"unster, Institute for Theoretical Physics, 48149 M\"unster, Germany}
\affiliation{Theoretical Physics Department, CERN, 1211 Geneva 23, Switzerland}

\begin{abstract}
Cosmic strings are predicted by many Standard Model extensions involving the cosmological breaking of a symmetry with nontrivial first homotopy group and represent a potential source of primordial gravitational waves (GWs).
Present efforts to model the GW signal from cosmic strings are often based on minimal models, such as, \eg, the Nambu--Goto action that describes cosmic strings as exactly one-dimensional objects without any internal structure.
In order to arrive at more realistic predictions, it is therefore necessary to consider nonminimal models that make an attempt at accounting for the microscopic properties of cosmic strings.
With this goal in mind, we derive in this paper the GW spectrum emitted by current-carrying cosmic strings (CCCSs), which may form in a variety of cosmological scenarios.
Our analysis is based on a generalized version of the velocity-dependent one-scale (VOS) model, which, in addition to the mean velocity and correlation length of the string network, also describes the evolution of a chiral (light-like) current.
As we are able to show, the solutions of the VOS equations imply a temporarily growing fractional cosmic-string energy density, $\Omega_{\rm cs}$.
This results in an enhanced GW signal across a broad frequency interval, whose boundaries are determined by the times of generation and decay of cosmic-string currents.
Our findings have important implications for GW experiments in the Hz to MHz band and motivate the construction of realistic particle physics models that give rise to large currents on cosmic strings.

\end{abstract}

\maketitle

\section{Introduction}
\label{sec:introduction}


Cosmic defects are a common prediction in many models of physics beyond the Standard Model (BSM), especially, in the context of grand unified theories (GUTs)~\cite{PhysRevLett.33.451,PhysRevLett.32.438}.
Cosmological phase transitions associated with the spontaneous breaking of new global or local symmetries can in particular give rise to various types of topological and nontopological defects~\cite{Kibble:1976sj,Kibble:1980mv}, including domain walls~\cite{PhysRevLett.48.1156}, cosmic strings~\cite{Hindmarsh:1994re}, and monopoles~\cite{PhysRevLett.43.1365}.
Evidence for any type of cosmic defect would clearly point to new physics, since the two phase transitions predicted by the Standard Model, the QCD crossover~\cite{PhysRevD.29.338,Aoki:2006we,PhysRevLett.65.2491} and the electroweak crossover~\cite{Kajantie:1996qd,Gurtler:1997hr,Laine:1998jb}, are both expected to yield no observational signal from cosmic defects.
In this paper, we are going to focus on cosmic strings, which, unlike monopoles and domain walls, do not threaten to overclose the Universe, as long as the string network reaches its characteristic scaling regime.


Cosmic strings contribute to the primordial scalar power spectrum and can hence be probed in observations of the cosmic microwave background~\cite{Planck:2013mgr}.
Another appealing avenue for the detection of cosmic strings is the search for primordial gravitational waves (GWs)~\cite{Auclair:2019wcv}.
The GW signal from a string network is dominated by the emission from closed string loops~\cite{PhysRevD.31.3052}, which are formed when long strings in the network intersect with each other or with themselves.
String loops oscillate under their own tension and hence emit GWs.
In addition, localized features on closed loops, such as cusps and kinks, emit bursts of GW radiation, which all together results in a stochastic GW background (SGWB) over a large frequency range.
An interesting property of this SGWB signal is that it acts as a logbook of the expansion history of the early Universe.
Indeed, the GW signal emitted by cosmic strings encodes information on the state of the Universe when cosmic string loops were formed and emitted GWs~\cite{Cui:2018rwi}.
%
This dependence on the expansion rate notably results in a flat GW spectrum across many orders of magnitude in frequency that is associated with the era of radiation domination in the early Universe.
Similarly, loops produced during the matter era result in a steeply falling contribution to the GW spectrum when going from higher to lower frequencies~\cite{Auclair:2019wcv}.
In this sense, cosmic strings can be used to probe nonstandard expansion histories of the early Universe, such as early matter domination~\cite{Gouttenoire:2019rtn,Blasi:2020wpy} and kination~\cite{Samanta:2021zzk,Gouttenoire:2021jhk}, see also Ref.~\cite{Gouttenoire:2019kij}.


\begin{figure}
\centering
\includegraphics[width=0.8\textwidth]{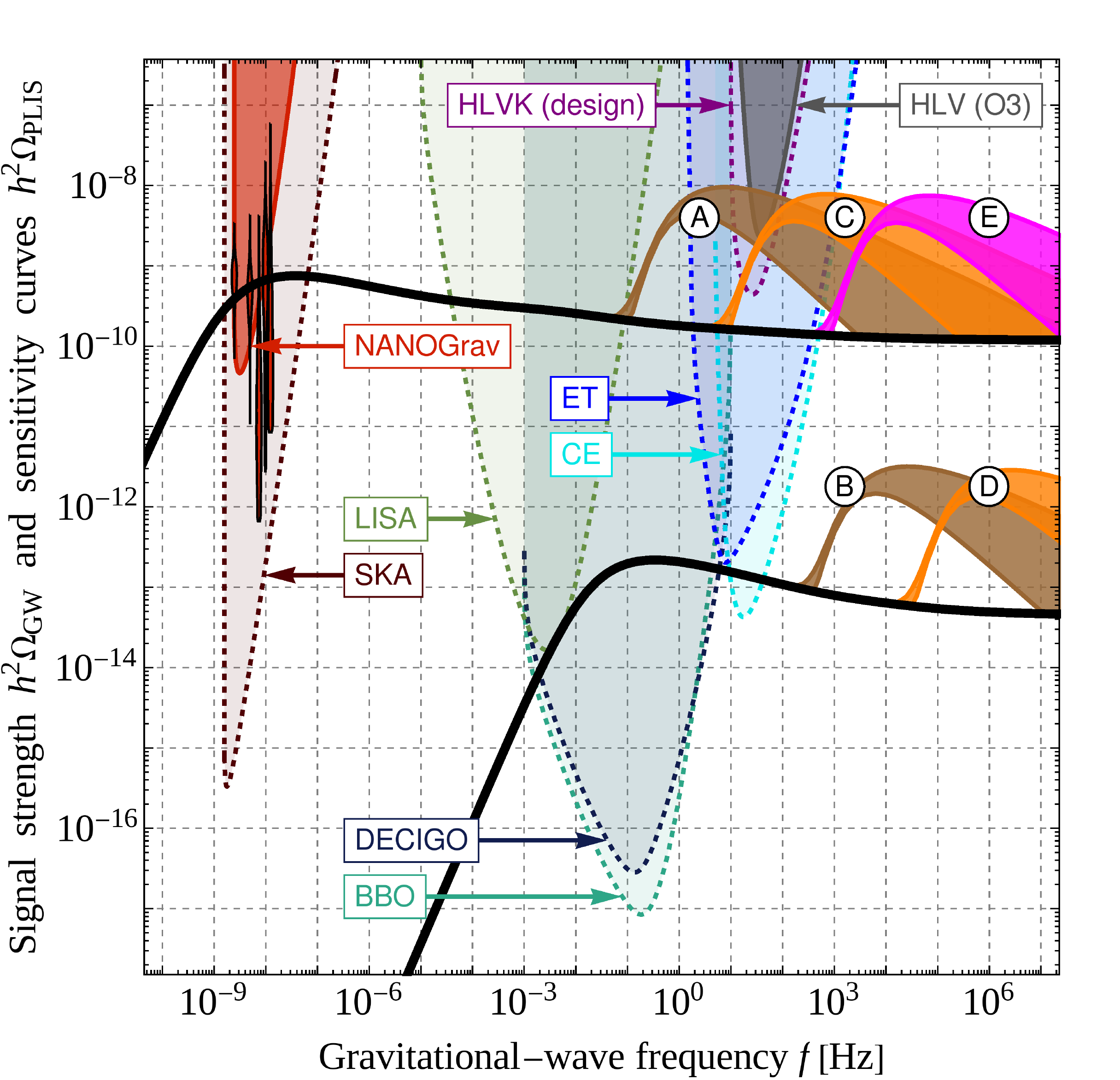}
\caption{Benchmark GW spectra emitted by current-carrying cosmic strings. For spectra (A), (C), (E), we assume a cosmic-string tension of $G\mu=0.5\times 10^{-10}$, while for spectra (B) and (E), we assume $G\mu=0.5\times 10^{-17}$. In scenarios (A) and (B), the current flowing on cosmic strings decays around the time of the QCD crossover, while in scenarios (C) and (D), it decays around the time of the electroweak crossover. In scenario (E), the current decays at even earlier times (see text for details). For each scenario, the lower and upper edges of the predicted spectrum respectively correspond to $k=1$ and $k=10^6$ harmonic string modes accounted for in the computation of the GW spectrum. We also show standard (\ie, no current) spectra for string networks with $G\mu=0.5\times 10^{-10}$ and $G\mu=0.5\times 10^{-17}$ (black lines), together with existing constraints and future sensitivities of present and planned GW experiments~\cite{Schmitz:2020syl}. All spectra for $G\mu=0.5\times 10^{-10}$ can explain the PTA signal at nHz frequencies.}
\label{fig:money}
\end{figure}


Recently, GWs from cosmic strings attracted a lot of attention in consequence of the first hints for a SGWB signal at nHz frequencies reported by the NANOGrav, PPTA, EPTA, and IPTA pulsar timing array (PTA) collaborations~\cite{NANOGrav:2020bcs,Goncharov:2021oub,Chen:2021rqp,Antoniadis:2022pcn}.
As demonstrated in Refs.~\cite{Blasi:2020mfx,Ellis:2020ena,Buchmuller:2020lbh,Bian:2022tju,Chen:2022azo}, GWs from cosmic strings represent a plausible interpretation of the PTA data and hence compete with the astrophysical interpretation in terms of inspiraling supermassive black-hole binaries~\cite{Middleton:2020asl}.
At present, the true nature of the new PTA signal is still unclear, as the characteristic interpulsar quadrupole correlations that would clearly indicate a GW origin have not yet been confirmed.
However, in anticipation of near-future PTA data, cosmic strings remain one of the leading candidates for the source of a SGWB at nHz frequencies, which calls for refined theoretical predictions of the expected GW signal.


The precise description of a string network requires large-scale numerical simulations~\cite{Albrecht:1989mk}, which are typically either based on the Nambu--Goto action~\cite{Ringeval:2005kr,Blanco-Pillado:2011egf,Blanco-Pillado:2013qja,Blanco-Pillado:2017oxo,Blanco-Pillado:2019tbi} or a field-theoretic description in terms of the Abelian Higgs model~\cite{Vincent:1997cx,Olum:1998ag,Hindmarsh:2008dw,Hindmarsh:2017qff,Matsunami:2019fss,Hindmarsh:2021mnl}; for a comparison of Abelian Higgs and Nambu--Goto strings, we refer to Sec.~3.4 of Ref.~\cite{Auclair:2019wcv}.
In the Nambu--Goto approximation, cosmic strings are treated as exactly one-dimensional featureless objects whose only relevant property is their tension or energy per unit length, $\mu$.
In order to obtain more realistic predictions, it is therefore necessary to go beyond this minimal description and consider, \eg, extensions of the simplest Nambu--Goto model featuring additional worldsheet degrees of freedom.
Such degrees of freedom can be currents induced by charge carriers that propagate along cosmic strings~\cite{Oliveira:2012nj,Martins:2020jbq}, or, \eg, short-wavelength propagation modes known as wiggles \cite{Almeida:2021ihc}.
In this paper, we will focus on the former possibility and consider \emph{neutral currents}, \ie, currents of particles that are only charged under the Abelian gauge group whose spontaneous breaking results in the formation of the string network but that do not interact with any other gauge force.
This notably ensures that strings cannot lose energy via their coupling to massless vector bosons, as would be the case if they carried electromagnetic currents and which would in fact lead to a highly suppressed GW emission~\cite{Ostriker:1986xc}.


Current-carrying cosmic strings (CCCSs) were first introduced by Witten~\cite{Witten:1984eb}, who studied both bosonic and fermionic charge carriers.
More recently, CCCSs were investigated in Refs.~\cite{Hartmann:2017lno,Brandenberger:2019lfm,Imtiaz:2020igv,Theriault:2021mrq,Martins:2021cid,Cyr:2022urs,Arias:2022fcd}.
The evolution of standard Nambu--Goto strings can be conveniently described by the so-called velocity-dependent one-scale (VOS) model \cite{Martins:1996jp,Martins:2000cs}, which allows one to track the evolution of two important properties of the string network: its correlation length and the root-mean-square velocity of long strings.
By comparison, a network of CCCSs features on top a third property that is not described by the VOS model: the strength of the current propagating on cosmic strings.
Recently, the authors of Ref.~\cite{Martins:2020jbq} generalized the standard VOS model in order to account for the time evolution of this current.
In our analysis, we will make use of these results and solve the generalized VOS equations derived in Ref.~\cite{Martins:2020jbq} for a CCCS network featuring a chiral (light-like) current.
This corresponds to charge carriers that are massive far away from the string core, but allow for massless zero modes along the string, where the symmetry is restored.
In combination with the fact that we assume neutral currents, a well-motivated CCCS scenario appears to be right-handed neutrinos (RHNs) propagating along cosmic $B\!-\!L$ cosmic strings, which can be easily embedded in grand unified theories (GUTs) based on $SO\left(10\right)$.

A first study of RHN scattering and capture by cosmic strings has been carried out in Ref.~\cite{Chavez:2002sm}.
On the other hand, superconducting currents made of zero modes can undergo a sudden change when these zero mode solutions are no longer permitted as a result of a phase transition.
This may \emph{e.g.} apply to RHN currents at the time of the electroweak phase transition, when the Higgs vacuum expectation value gives mass to RHN zero modes propagating along the string cores~\cite{Davis:1996sp}.

%
%
As we will see below, the decay of RHN currents at the time of electroweak symmetry breaking can then be instrumental in avoiding a potential overclosure problem that may arise if the currents carried by the string network are too long-lived.
Overall, we, however, emphasize that the process of current generation and quenching requires further work, especially, in the context of specific GUT models.
In this paper, we will not address this question and defer any further top-down model building to future work.
Instead, we will adopt a bottom-up approach and carry out a phenomenological analysis based on the recently proposed generalized VOS model focusing on a general scenario where currents are present only for a certain amount of time between the instant of network formation and today.
As can be seen in \cref{fig:money}, this approach leads to very promising results, in particular, a boosted SGWB signal over large ranges in frequency that is related to a temporary growth in the fractional CCCS energy density, $\Omega_{\rm cs}$.
We therefore hope that our analysis is going to motivate further model-building efforts aiming at the construction of realistic particle physics models that give rise to large currents on cosmic strings.


The remainder of this paper is organized as follows.
First, we will discuss the generalized VOS model in \cref{sec:vos}.
We will start by introducing the generalized VOS equations in \cref{sec:eqs}, which include the new terms accounting for the presence of a current.
We will then derive the attractor solutions of these equations, first in the case of zero current in \cref{sec:vanish} and subsequently in the case of nonvanishing current in \cref{sec:steady}, which is the scenario of our interest.
In \cref{sec:loop_number}, we will employ these results in order to derive the loop number density as well as the energy density of a CCCS network.
The SGWB signal is computed in \cref{sec:gw}, where we compare our predictions to the existing and projected sensitivities of current and planned GW experiments.
\cref{sec:conclusions}, finally, contains our conclusions.


\section{Generalized VOS model with chiral currents}
\label{sec:vos}


\noindent
The energy density of a network of long strings is
\begin{equation}
	\rho_\infty = \frac{\mu}{L^2}\,,
	\label{eq:rho_infty}
\end{equation}
where $L$ is the characteristic correlation length.
A network of Nambu--Goto strings in the expanding Universe with no worldsheet degrees of freedom and intercommutation probability $P=1$ is known to reach a scaling solution, $L\propto t$, soon after formation.
The existence of such a solution can be derived analytically within the VOS framework~\cite{Martins:1996jp,Martins:2000cs,Auclair:2019wcv}, which in its standard form contains two coupled differential equations describing the evolution of $L$ and the root-mean-square (RMS) velocity $v$.


\subsection{Nonlinear autonomous system}
\label{sec:eqs}

Using $t$ to denote cosmic time, the generalized VOS equations for superconducting strings carrying a chiral current read \cite{Martins:2020jbq}\footnote{Note that we express these equations in terms of cosmic time and physical length whereas in Ref.~\cite{Martins:2020jbq} conformal time and wavenumber are used.}
\begin{subequations}
	\begin{align} \label{eq:master}
		\dot{L} & = \frac{\dot{a}}{a} \frac{L}{1+Y}\qty(1+v^{2}+2Y) + \frac{g \tilde{c}}{2\sqrt{1+Y}}v\,,                           \\
		\dot{v} & = \frac{1-v^{2}}{1+Y} \left[\frac{(1-Y) k(v)}{L \sqrt{1+Y}} - 2 v \frac{\dot{a}}{a}\right]\,,                 \\
		\dot{Y} & = 2Y \left[\frac{v k(v)}{L\sqrt{1+Y}} - \frac{\dot{a}}{a} \right] - \frac{v}{L} \tilde{c} (g-1) \sqrt{1+Y}\,, \\
		k(v)    & = \frac{2\sqrt{2}}{\pi} \frac{1-8v^{6}}{1+8v^{6}}\qty(1-v^{2})\qty(1+2\sqrt{2}v^{3})\,. \label{eq:master-k}
	\end{align}
\end{subequations}
Here, $Y$ denotes the current strength defined as $Y=(Q^2+J^2)/2$, where $Q$ and $J$
are the total charge and current energy density of the system, respectively.
The three quantities $Y$, $Q^2$, and $J^2$ are dimensionless and understood to describe the properties of the current carried by the string network in units of the string tension $\mu$.
Chiral (or light-like) currents are characterized by a vanishing state parameter $K = Q^2 - J^2$, meaning that in our scenario of interest $K= 0$ and hence $Y = Q^2 = J^2$.
Further, $a$ in the above set of equations denotes the cosmic scale factor.
$\tilde{c}$ is the so-called chopping parameter, which describes the efficiency of chopping off loops from the long-string network.
In absence of dedicated numerical simulations of CCCS networks, we fix $\tilde{c}\approx 0.23$, the typical value found in numerical simulations of standard uncharged Nambu--Goto strings~\cite{Martins:2000cs}.
We also set $g=\sqrt{1+Y}$, in which case the energy loss of the long-string network due to loop formation is qualitatively described in the same way as in the current-less case; see Eqs.~(33), (38), and (39) in Ref.~\cite{Martins:2020jbq}.
Finally, the function $k(v)$ is known as the momentum parameter; again we adopt the standard expression for uncharged Nambu-Goto strings \cite{Auclair:2019wcv,Martins:2020jbq}.

\begin{figure}
	\includegraphics[width=\textwidth]{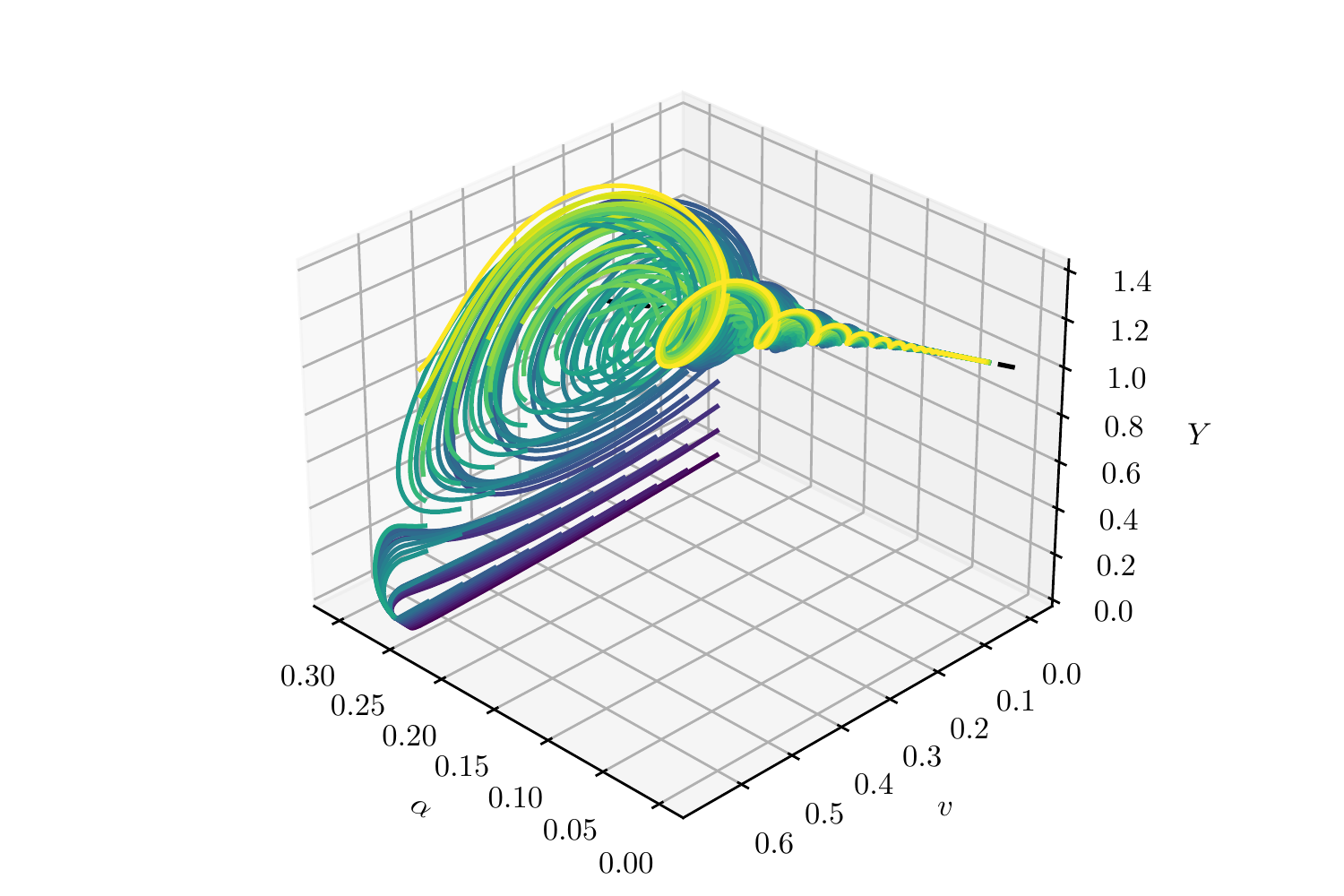}
	\caption{Streamlines derived from the autonomous system of ordinary differential equations \eqref{eq:autonomous-alpha}, \eqref{eq:autonomous-v} and \eqref{eq:autonomous-y} for $\tilde{c}=0.23$ and $\nu=1/2$. All the trajectories start from the plane $\alpha = 0.27$ and eventually fall into one of the two different attractors, one at $Y=0$ and another one at $Y\sim 1$.}
	\label{fig:streamtube}
\end{figure}

This system of ordinary differential equations can be rewritten in terms of the scaling length $\alpha(t) = L(t) / t$ and derivatives with respect to logarithmic time
\begin{subequations}
	\begin{align}
		\dv{\alpha}{\ln t} & = \frac{\alpha\nu}{1+Y}\qty(1+v^{2}+2Y) - \alpha + \frac{g \tilde{c}}{2\sqrt{1+Y}}v\,,	\label{eq:autonomous-alpha}                     \\
		\dv{v}{\ln t}      & = \frac{1-v^{2}}{1+Y} \left[\frac{(1-Y) k(v)}{\alpha \sqrt{1+Y}} - 2 v \nu\right]\,,     	\label{eq:autonomous-v}                  \\
		\dv{Y}{\ln t}      & = 2Y \left[\frac{v k(v)}{\alpha\sqrt{1+Y}} - \nu \right] - \frac{v}{\alpha} \tilde{c} (g-1) \sqrt{1+Y}\,,	\label{eq:autonomous-y}
	\end{align}
\end{subequations}
where $\nu=1/2$ ($\nu=2/3$) for radiation (matter) dominated Universe since we defined $a(t)\sim t^\nu$.
Notice that the right-hand side does not depend explicitly on time, meaning that this set of equations is an \emph{autonomous system} and the vector $X(\ln t) = (\alpha, v, Y)(\ln t)$ describing the solution of this system is an \emph{integral curve}.
We showcase in \cref{fig:streamtube} the integral curves for $X(\ln t)$ in three-dimensional space.

Looking at \cref{fig:streamtube}, we identify two attractor solutions.
First, there is an attractor solution where the current vanishes and the parameters $\alpha$ and $v$ reach a constant value, identical to the predictions of the standard VOS framework.
We discuss it in more details in \cref{sec:vanish}.
A second attractor solution exists when $Y$ is large enough, in which case the current remains close to $Y=1$ and the parameters $\alpha$ and $v$ feature a power-law decay.
We cover this situation in \cref{sec:steady}.

\subsection{Vanishing current attractor}
\label{sec:vanish}

In this section, we study the autonomous system of Eqs.~\eqref{eq:autonomous-alpha}, \eqref{eq:autonomous-v} and \eqref{eq:autonomous-y} around the attractor appearing at $Y=0$.
In what follows we will first determine the exact location of the attractor, \ie the fixed point for this system of equations around $Y=0$. Then we will determine its characteristics, such as its stability and the efficiency with which the system converges to $Y= 0$.

\subsubsection{Fixed point at vanishing current}
\label{sec:vanish-fix}

First, we note that setting $Y=0$ trivially satisfies Eq. \eqref{eq:autonomous-y}.
Then, imposing that $\dv*{\alpha}{\ln t} = 0$, we obtain that
\begin{equation}
	\alpha = \frac{\tilde{c} v_\infty}{2(1-\nu-\nu v_\infty^2)}\,,
\end{equation}
where $v_\infty$ is the RMS velocity at the fixed point.
Further, if we impose that $\dv*{v}{\ln t} = 0$, we find that the velocity has to satisfy the polynomial equation
\begin{equation}
	k(v_\infty) = \frac{\tilde{c} \nu v_\infty^2}{1 - \nu - \nu v_\infty^2}\,,
	\label{eq:k}
\end{equation}
where $k(v)$ is defined in Eq. \eqref{eq:master-k}; the solution of \cref{eq:k} cannot be determined analytically.
We provide numerical values for $v_\infty$ in \cref{table:vanish}.

\subsubsection{Characteristics of the $Y=0$ attractor solution}
\label{sec:vanish-attractor}

One can assess the behaviour of $Y$ around the $Y=0$ fixed point as follows:
Let us divide Eq.~\eqref{eq:autonomous-y} by $Y$, then expand for $Y \rightarrow 0$ and evaluate at the fixed point. We thus find a power-law behavior $Y \propto t^\beta$, with the coefficient $\beta$ determined by
\begin{equation}
	\beta \equiv \frac{1}{Y} \dv{Y}{\ln t} = \frac{2 k(v_\infty) v_\infty}{\alpha} - \frac{c v_\infty}{2 \alpha} - 2\nu = -1 + (5 v_\infty^2 - 1) \nu\,.
	\label{eq:beta}
\end{equation}
Numerical values for $\beta$ are also given in \cref{table:vanish}.

In order to prove that this fixed point is indeed an attractor we utilize that
 the system of Eqs. \eqref{eq:autonomous-alpha}, \eqref{eq:autonomous-v} and \eqref{eq:autonomous-y} is an autonomous system of the form
\begin{equation}
	\dot{X} = f(X)\,,
\end{equation}
with a fixed point $X_p = (\alpha, v_\infty, 0)$, \ie $f(X_p) = 0$. The Jacobian of $f(X)$ evaluated at $X_p$ reads
\begin{equation}
	\begin{pmatrix}
		-1 + \nu (1 + v_\infty^2)                                          & \tilde{c} \qty[\frac{1}{2} + \frac{\nu v_\infty^2}{1 - \nu (1+v_\infty^2)}] & \frac{\tilde{c} \nu v_\infty}{2 [1 + \nu(1+v_\infty^2)]}(1-v_\infty^2) \\
		- \frac{4 \nu}{\tilde{c}} (1 - v_\infty^2)[1 - \nu (1+v_\infty^2)] & - 2 \nu (1 - v_\infty^2)                                                    & -3 \nu v_\infty ( 1 - v_\infty^2)                                      \\
		0                                                                  & 0                                                                           & -1 +\nu (5v_\infty^2 - 1).
	\end{pmatrix}\,.
\end{equation}
Looking at the last row of the Jacobian, we identify one of its eigenvalues, dubbed $\lambda_3 = -1 +\nu (5v_\infty^2 - 1)$ which matches $\beta$ from Eq. \eqref{eq:beta}.
For the fixed point to be an attractor, the three eigenvalues $\lambda_1, \lambda_2$ and $\lambda_3$ are required to have a negative real part.
From the trace and determinant of the Jacobian, we find
\begin{subequations}
	\begin{align}
		\lambda_1 + \lambda_2      & = -1 +\nu (3 v_\infty^2 - 1) \in \mathbb{R}\,, \\
		\lambda_1 \cdot \lambda_2 & = 4 \nu(1 - v_\infty^2) (1 - \nu) > 0 \,.
	\end{align}
\end{subequations}
Hence, $\lambda_1$ and $\lambda_2$ are complex conjugate and
\begin{equation}
	\Re{\lambda_1} = \Re{\lambda_2} = \frac{1}{2} \qty[-1 +\nu (3 v_\infty^2 - 1)]\,.
\end{equation}
Numerical values for $\Re{\lambda_1}$ are shown in \cref{table:vanish} for several cases; for all values of $\tilde{c}$ we have considered, the fixed point $Y=0$ is a stable attractor.

\begin{table}[t!]
	\centering
	\setlength{\tabcolsep}{16pt}
	\begin{tabular}{ccccc}
		$\tilde{c}$ & $\alpha$ & $v_{\infty}$ & $\beta = \lambda_3$ & $\Re{\lambda_1} = \Re{\lambda_2}$ \\
		\hline
		\hline
		$0.15$      & $0.186$  & $0.675$      & $- 0.360$           & $- 0.408$                         \\
		$0.20$      & $0.24$   & $0.667$      & $- 0.388$           & $- 0.417$                         \\
		$0.23$      & $0.271$  & $0.662$      & $- 0.404$           & $- 0.421$                         \\
		$0.25$      & $0.291$  & $0.659$      & $- 0.414$           & $- 0.424$                         \\
		$0.30$      & $0.340$  & $0.652$      & $- 0.438$           & $- 0.431$                         \\
		\hline
		\hline
	\end{tabular}
	\caption{Values of parameters $\alpha$, $v_\infty$ and $\beta$ around the vanishing current attractor of \cref{sec:vanish} for $\nu = 1/2$.}
	\label{table:vanish}
\end{table}

\subsection{Steady current attractor}
\label{sec:steady}

\begin{figure}
	\includegraphics[width=.49\textwidth]{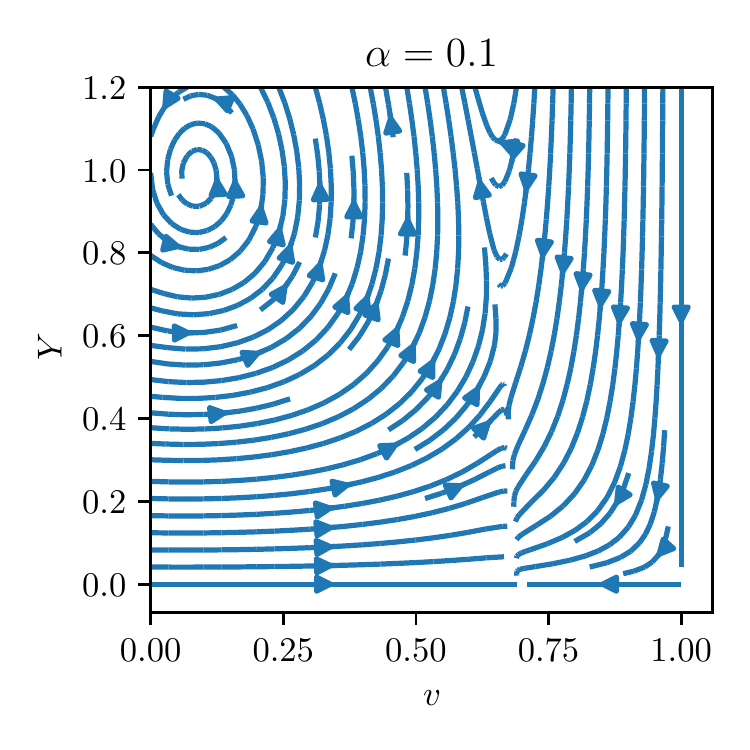}
	\includegraphics[width=.49\textwidth]{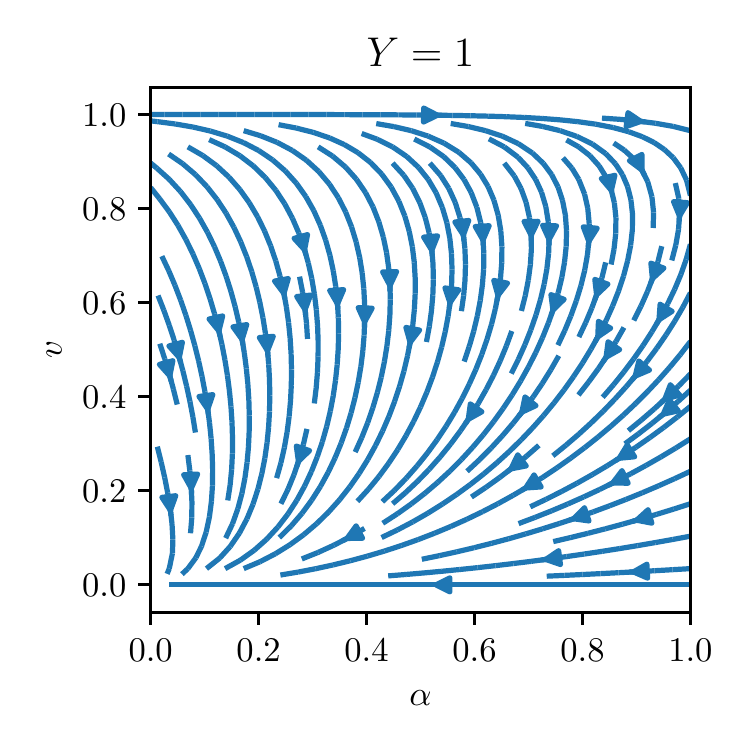}
	\caption{Two-dimensional slices of \cref{fig:streamtube} around the $Y\sim 1$ attractor.}
	\label{fig:streamplots}
\end{figure}

Now we focus our attention on the attractor solution around $Y=1$ identified in \cref{fig:streamtube}.
In \cref{fig:streamplots}, we show slices of this streamplot across the planes $\alpha = 0.1$ and $Y=1$. Clearly, around this attractor solution, the parameters $\alpha$, $v$ and $\epsilon \equiv 1 - Y$ converge to $0$.
In what follows, we describe the attractor solution and discuss its stability.

\subsubsection{Description of the attractor solution}
\label{sec:steady-attractor}

We start again from Eqs. \eqref{eq:autonomous-alpha}, \eqref{eq:autonomous-v} and \eqref{eq:autonomous-y} and expand around $Y=1$, such that $\abs{\epsilon(t)} \ll 1$.
We obtain the following \emph{nonlinear autonomous system} of differential equations
\begin{subequations}
	\begin{align}
		\dv{\alpha}{\ln t}     & = \nu \frac{\alpha }{2}\qty(3+v^{2} - \frac{2}{\nu}) + \frac{\tilde{c}}{2}v\,,                     \\
		\dv{v}{\ln t}          & = \frac{1-v^{2}}{2} \qty[\frac{\epsilon k(v)}{\alpha(t) \sqrt{2}} - 2 v \nu]\,,                     \\
		- \dv{\epsilon}{\ln t} & = 2 \qty[\frac{v k(v)}{\alpha \sqrt{2}} - \nu ] - \frac{v}{\alpha} \tilde{c} \qty(2 - \sqrt{2})\,.
	\end{align}
\end{subequations}
We make the additional assumption that  $\alpha(t) \ll 1$ and $v(t) \ll 1$ after a certain amount of time, thus reducing the system of differential equations to
\begin{subequations}
	\begin{align}
		\dv{\alpha}{\ln t}   & = \frac{3\nu\alpha}{2} - \alpha + \frac{\tilde{c}}{2} v\,,    \label{eq:steady-alpha}                         \\
		\dv{v}{\ln t}        & = \frac{\epsilon}{\alpha \pi} - \nu v\,,  \label{eq:steady-velocity}                                          \\
		\dv{\epsilon}{\ln t} & = 2\nu + \frac{v}{\alpha} \tilde{c} \qty(2 - \sqrt{2} - \frac{4}{\tilde{c}\pi})\,. \label{eq:steady-epsilon}
	\end{align}
\end{subequations}
One should note that even though we numerically observe that $(\alpha, v, \epsilon) = (0, 0, 0)$ is an attractor in our dynamical system, it is not a fixed point.
Indeed, Eqs. \eqref{eq:steady-velocity} and \eqref{eq:steady-epsilon} are singular when $\alpha = 0$.
In the following, we determine the attractor trajectory when $\alpha(t) \rightarrow 0$, such that Eqs. \eqref{eq:steady-alpha}, \eqref{eq:steady-velocity} and \eqref{eq:steady-epsilon} converge to $0$.
Imposing $\abs{\dv*{\epsilon}{\ln t}} \ll 1$, we find that $\alpha$ and $v$ are proportional to each other
\begin{equation}
	\kappa \equiv \frac{v}{\alpha} = -\frac{2\nu}{\tilde{c}\qty(2-\sqrt{2})-4 / \pi}\,.
	\label{eq:voveralpha}
\end{equation}
At this point, if we impose that $\abs{\dv*{\alpha}{\ln t}} \ll 1$, we recover $\abs{\alpha} \ll 1$.
However, we can go further and show that $\alpha$ follows a power-law decay, $\alpha \propto t^\delta$, around the attractor with coefficient
\begin{equation}
	\delta \equiv \frac{1}{\alpha}\dv{\alpha}{\ln t} = \frac{3\nu}{2}-1 - \frac{\nu}{\qty(2-\sqrt{2})-4 / (\tilde{c}\pi)}\,.
\end{equation}

\begin{figure}
	\centering
	\includegraphics{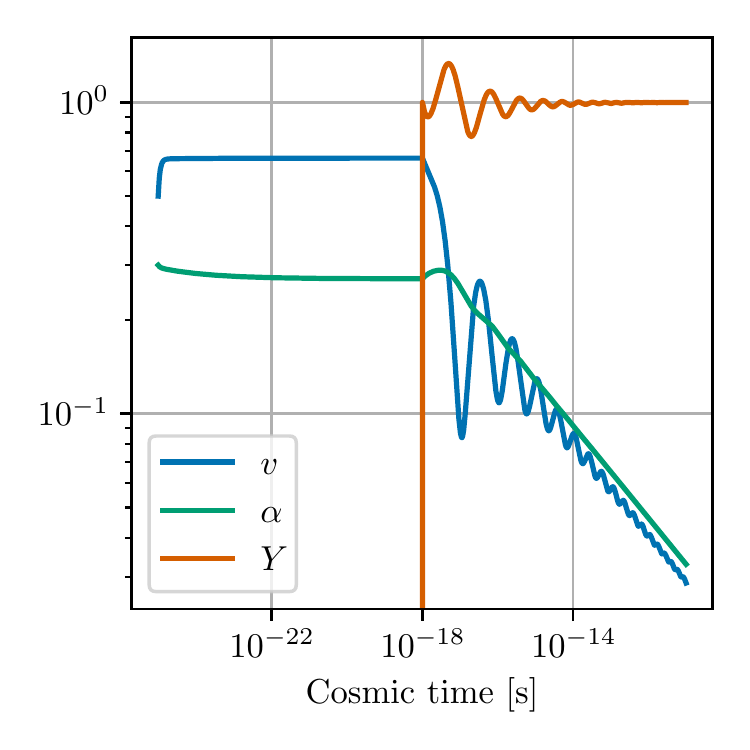}
	\includegraphics{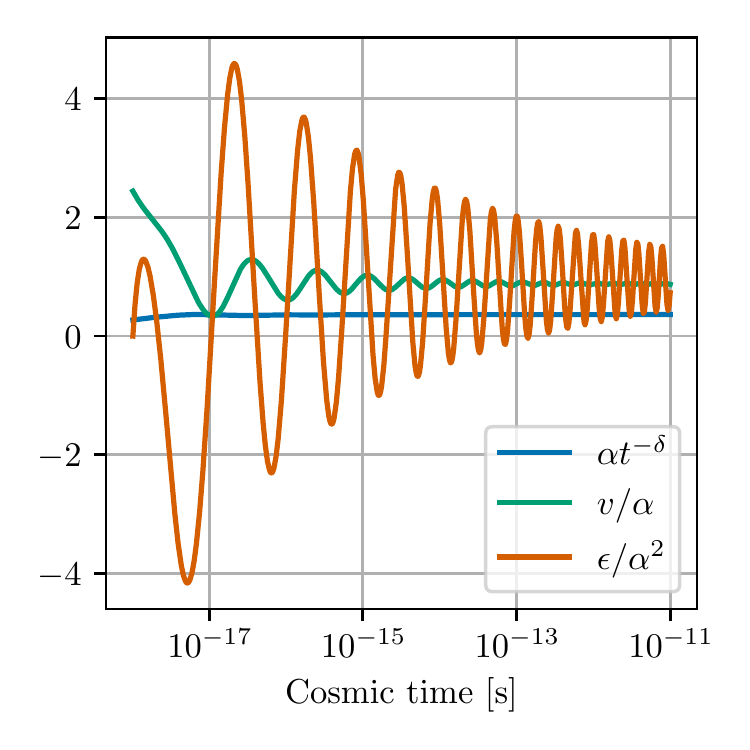}
	\caption{\emph{Left panel}: Evolution of the parameters $\alpha$, $v_\infty$ and $Y$ when the current is artificially sourced to $1$ at $t=10^{-18}$ s, for $\tilde{c} = 0.23$ and $\nu=1/2$. \emph{Right panel}: Evolution of $\alpha t^{-\delta}$, $v/\alpha$ and $\epsilon/\alpha^2$ that should be approximately time independent for a steady current attractor.}
	\label{fig:transition}
\end{figure}

Similarly, if we divide \cref{eq:steady-velocity} by $v$, considering that $v$ is proportional to $\alpha$, we obtain that
\begin{equation}
	\delta = \frac{1}{v} \dv{v}{\ln t} = \frac{\epsilon}{\alpha v \pi} - \nu\,.
\end{equation}
Thus, we deduce that $\epsilon$ is proportional to $\alpha^2$
\begin{equation}
	\frac{\epsilon}{\alpha^2} =(\delta+\nu) \pi \frac{v}{\alpha}\,.
	\label{eq:epsilonoveralpha}
\end{equation}
The numerical values for all these coefficients are in \cref{table:steady}.
In the left panel of \cref{fig:transition} we show the evolution of $\alpha$, $v$ and $Y$ in the scenario where the current suddenly increases from $Y=0$ to $Y=1$ at a particular time.
Given $\alpha \propto t^\delta$ and derived \cref{eq:voveralpha,eq:epsilonoveralpha}, in the right panel of
\cref{fig:transition} we show the evolution of expressions that should be approximately time-independent for steady current attractor.

\begin{table}[t!]
	\centering
	\setlength{\tabcolsep}{16pt}
	\begin{tabular}{ccccc}
		$\tilde{c}$ & $\kappa \equiv v_\infty / \alpha$ & $\epsilon / \alpha ^2$ & $\delta$      & $Y_\mathrm{min}$ \\
		\hline
		\hline
		$0.15$      & $0.844$             & $0.830$                & $-0.187$ & $0.36$           \\
		$0.20$      & $0.865$             & $0.914$                & $-0.164$ & $0.42$           \\
		$0.23$      & $0.878$             & $0.969$                & $-0.149$ & $0.46$           \\
		$0.25$      & $0.887$             & $1.001$                & $-0.139$ & $0.49$           \\
		$0.30$      & $0.911$             & $1.107$                & $-0.113$ & $0.58$           \\
		\hline
		\hline
	\end{tabular}
	\caption{Values of parameters $v_\infty / \alpha$ and $\delta$ determining the steady current attractor of \cref{sec:steady} for $\nu=1/2$.}
	\label{table:steady}
\end{table}

\subsubsection{Stability of the attractor solution}
\label{sec:steady-stability}

Unlike in the case of the $Y=0$ attractor, the system of differential equations is not well defined at the steady current attractor $(\alpha, v, Y) = (0, 0, 1)$, and we cannot  straightforwardly perform the same stability analysis.
First, the existence of the steady current attractor requires that $\delta < 0$ which imposes
\begin{equation}
	\nu < \frac{8 - 2 ( 2 - \sqrt{2}) \tilde{c} \pi}{12 + (3 \sqrt{2} - 4) \tilde{c} \pi}\,.
	\label{eq:nu-max}
\end{equation}
If $\tilde{c}=0.23$, then $\nu < 0.588$, meaning that this attractor does not exist in the matter era where $\nu=2/3$.
Second, we estimate numerically the basin of attraction of the steady current attractor.
Assuming that the system starts from the vanishing current attractor and the current is instantaneously switched on, we compute the minimal value of the current, $Y_\mathrm{min}$, needed in order for the system to fall into the steady current attractor. Numerical values for $Y_\mathrm{min}$ are collected in \cref{table:steady}.
Our detailed analysis in this section is supported by the previous findings in the literature. While the behaviour of the vanishing current attractor solution follows from the standard VOS model, the behavior of the steady current solution has previously been discussed in Ref.~\cite{Oliveira:2012nj} (see also \cite{Martins:2021cid}); our findings are consistent with those in Ref.~\cite{Oliveira:2012nj}.

For novel gravitational wave signature from current-carrying cosmic strings  we require that the system approaches the steady current solution, otherwise the gravitational wave spectrum would match the standard one, see e.g. \cite{Cui:2018rwi}. We should also emphasize that the steady current solution $Y=1$
is near the edge of validity of the studied system of equations.

\section{Loop number density}
\label{sec:loop_number}
\noindent
Let us now derive the number density of cosmic string loops for the steady current attractor scenario, discussed in \cref{sec:steady}. This will allow us to calculate the gravitational-wave spectra (see \cref{sec:gw}).

\subsection{General expressions}

The length, $L(t)$, defined in Eq. \eqref{eq:rho_infty}, is directly related to the energy density of the long-string network.
The energy lost by the long string reads
\begin{equation}
	\dv{\rho_{\infty}}{t}
	= \frac{-2 \mu}{L^{3}} \dv{L}{t}
	=  \frac{-2 \mu}{L^{3}} \qty[ \frac{\dot{a}}{a} \frac{L}{1+Y}(1+v^{2}+2Y) + \frac{\tilde{c} v}{2} ] \,.
	\label{eq:energy-lost}
\end{equation}
The chopping of loops, driven by the last term of \cref{eq:energy-lost}, induces the following transfer of energy density to the cosmic string loops
\begin{equation}
	\dot{\rho}_{\infty, \,\text{loops}} = - \frac{\tilde{c}\mu v(t)}{L^3(t)}\,.
	\label{eq:loss}
\end{equation}

Assuming that a fraction, $\mathcal{F}$, of this energy goes
into large loops with boost factor $\gamma$ yields
\begin{equation}
    \mathcal{F} \dot{\rho}_{\infty, \,\mathrm{loops}}
    = - \gamma \mu \int \dd{\ell} \ell \mathcal{P}(\ell, t)\,.
    \label{eq:transfer-pierre}
\end{equation}
Here, the loop production function $\mathcal{P}(\ell, t)$ is the number of loops of sizes in the range $(\ell, \ell +\dd{\ell})$ produced in the time interval $(t, t+\dd{t})$.
Furthermore, we assume that all the loops are produced at the same scale with a fraction $\alpha_L$ of $L(t)$.
The loop production function then reads
\begin{equation}\label{eq:P}
    \mathcal{P}(\ell, t) = \frac{\tilde{c} \mathcal{F} v(t)}{\gamma \alpha_{L}\alpha^{4}(t)} t^{-5} \delta_D\qty[\alpha_L \alpha(t) - \frac{\ell}{t}],
\end{equation}
where $\delta_D$ is the Dirac delta distribution.

The loop number density is obtained by integrating the loop production function over the possible loop formation times
\begin{equation}
    n(\ell, t) = \frac{1}{a^{3}(t)} \int_{t_{\mathrm{ini}}}^{t} \dd{t'} a^{3}(t') \, \mathcal{P}(\ell'(t'), t')\,,
\end{equation}
where $\ell'(t') = \ell + \Gamma G\mu (t-t')$ is the hypothetical length of the loop at the time of formation.
The delta distribution in $\mathcal{P}$ imposes a unique choice for the loop formation time $t_\star$ satisfying
\begin{equation}
    \alpha_L \alpha(t_{\star})t_{\star} + \Gamma G\mu t_{\star} = \ell + \Gamma G\mu t\,.
    \label{eq:tstar}
\end{equation}

The loop number density is then given by
\begin{equation}\label{eq:t4n}
    t^{4} n(\ell, t) =  \frac{\tilde{c} \mathcal{F} v(t_\star)}{\gamma \alpha_{L}\alpha^{4}(t_\star)} \frac{\Theta(t_{\star} - t_{\mathrm{ini}}) \Theta(t - t_{\star})}{\Gamma G\mu + \alpha_L \alpha(t_{\star}) + \alpha_L \alpha'(t_{\star}) t_{\star}} \qty[\frac{a(t_{\star})}{a(t)}]^{3} \qty(\frac{t}{t_{\star}})^{4}\,,
\end{equation}
where $t_{\mathrm{ini}}$ is the time when the current is switched on.
Let us stress that there does not exist a closed form for $t_\star$ and we evaluate \cref{eq:tstar} numerically for all practical purposes. We, however, point out the existence of an analytical approximation (see \cref{sec:approx}) for the loop number density that works rather well and which helps in gaining an analytical understanding.

\subsection{Analytical approximations}
\label{sec:approx}

The equation for the loop number density in Eq.\,\eqref{eq:t4n}
can be simplified if we consider the limit in which $\alpha_L \alpha(t)\gg \Gamma G\mu$
for all the duration of the current-carrying phase.
This approximation allows us to solve
for the loop formation time, $t_\star$,
in a closed form
\begin{equation}\label{eq:appr}
    \ell + \Gamma G\mu t =
    [\alpha_L \alpha(t_\star)+\Gamma G \mu]\, t_\star
    \quad
    \Longrightarrow
    \quad
    t_\star \simeq t_{\mathrm{ini}} \left(\frac{\ell+\Gamma G\mu t}{\alpha_L \alpha_s t_{\mathrm{ini}}}\right)^{\frac{1}{1+\delta}},
\end{equation}
where we employed $\alpha(t)=\alpha_s \, (t/t_{\mathrm{ini}})^\delta$ parametrization according
to the asymptotic power--law behaviour described in \cref{sec:steady}; here $\alpha_s=\alpha(t_{\mathrm{ini}})$.

We then obtain
\begin{equation}
    t^4 n(\ell,t) \simeq
    \frac{\tilde c \mathcal{F} \kappa
    \alpha_L^2}{(1+\delta)\gamma}
      \frac{a(t_\star)^3}{a(t)^3}
    \left(x + \Gamma G\mu\right)^{-4}\Theta(t_{\star}-t_{\mathrm{ini}})
    \Theta(t - t_{\star})\,,
    \label{eq:lnd}
\end{equation}
where we have defined $x=\ell/t$, and $\kappa \equiv v(t_\star)/\alpha(t_\star)$
is an $\mathcal{O}(1)$ number within the current-carrying attractor solution, as shown in \cref{table:steady}.

The expression above can be further simplified if the time
at which the loop number density is evaluated and the loop formation time belong to the same cosmological era described by
$a(t)\propto t^\nu$. Then we obtain
\begin{equation}\label{eq:t4nfinal}
    t^4 n(\ell,t) \simeq
    \frac{ \tilde c \mathcal{F} \kappa \alpha_L^2}
    {(1+\delta)\gamma(\alpha_L \alpha_s)^{\frac{3\nu}{1+\delta}}}
    \qty(\frac{t}{t_{\mathrm{ini}}})^{-\frac{3\nu \delta}{1+\delta}}
    \left(x + \Gamma G\mu\right)^{-4+\frac{3\nu}{1+\delta}}\,
    \Theta(t_{\star}-t_{\mathrm{ini}})
    \Theta(t - t_{\star})\,.
\end{equation}

A couple of comments are in order. First, if we set $\delta=0$
we see that $n(\ell,t) t^4$ agrees with the usual scaling in which
$\ell$ and $t$ only enter through the dimensionless ratio $x$.
We then recover $n(\ell,t)t^4 \propto (x + \Gamma G \mu)^\beta$
with $\beta = -5/2 \, (-2)$ in radiation (matter) domination,
see \eg Ref.~\cite{Auclair:2019wcv}.

On the other hand, the loop number density does not scale
for $\delta \neq 0$, namely we still have an explicit dependence
on $t$ due to the presence of  $t_{\mathrm{ini}}$.
In particular, since $\delta \in (-1, 0)$
the loop number density increases with the
duration of the current-carrying phase, $t/t_{\mathrm{ini}}$.

\begin{figure}
	\centering
	\includegraphics[width=0.49\textwidth]{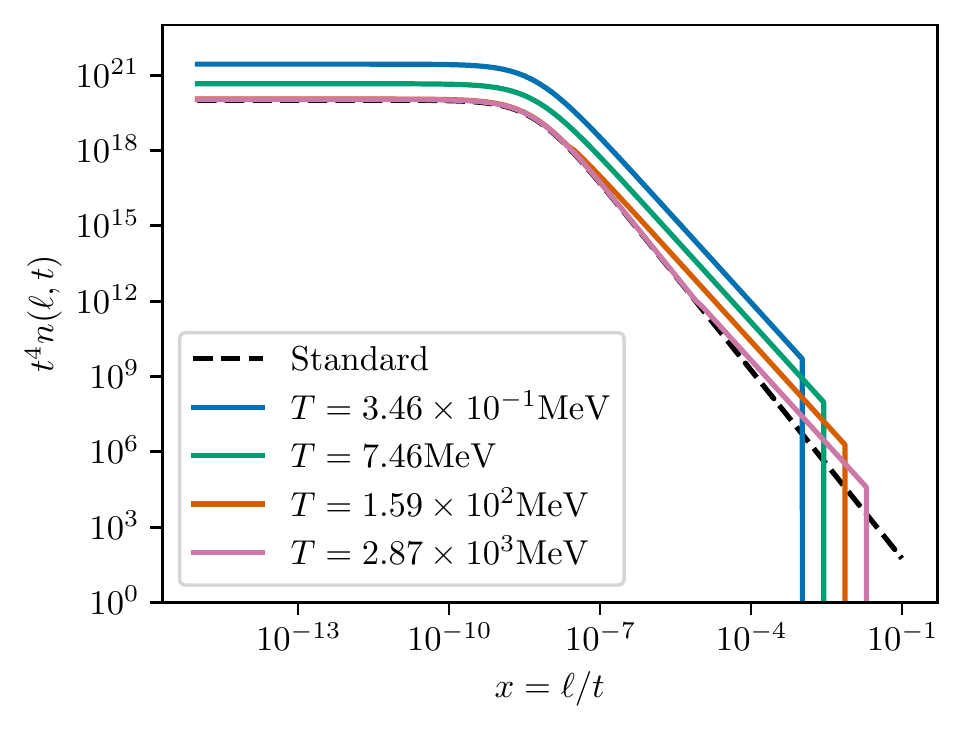}
	\includegraphics[width=0.49\textwidth]{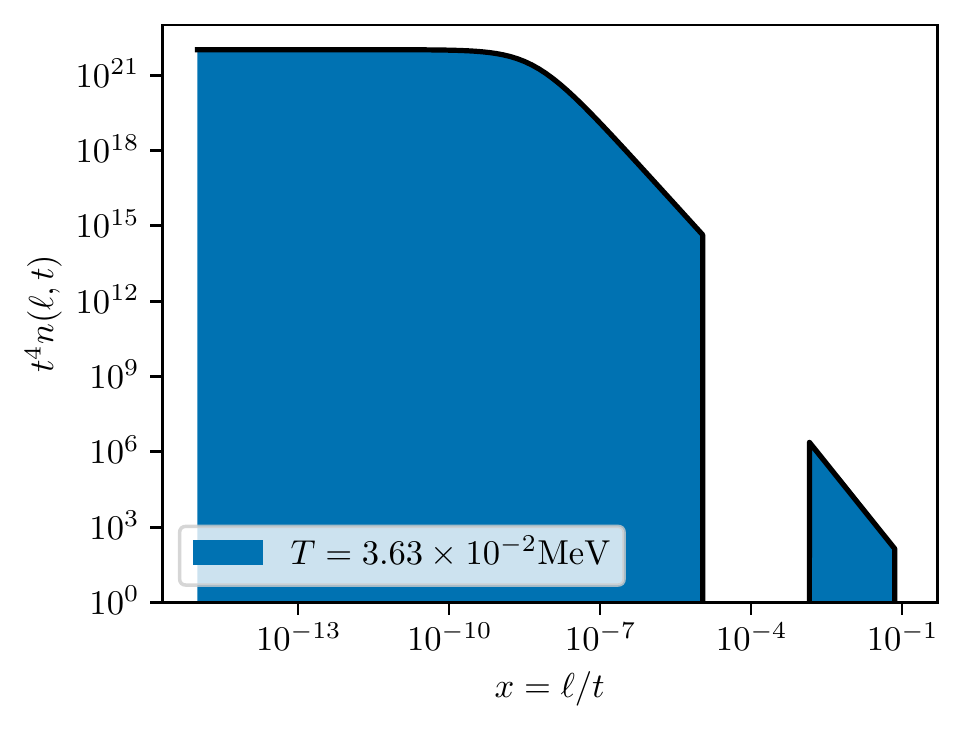}
	\caption{$t^4 n(\ell,t)$ for different times $t$ as a function
	of $x=\ell/t$. \emph{Left panel}: The black dashed line corresponds
	to the standard scaling. It can be seen that during the current-carrying phase more loops are produced at shorter lengths. In the calculation, we have taken $G\mu=10^{-10}$ and assumed that the transition starts at redshift $z=10^{15}$ and ends at $z=10^{9}$.
	\emph{Right panel}: $t^4 n(\ell,t)$ after the end of the current-carrying phase is shown. A gap develops in the loop distribution, see text for details.}
	\label{fig:n}
\end{figure}

We present the loop number density in the current-carrying
phase in \cref{fig:n} (left panel). Comparing to the standard case, we can infer that
 during the current-carrying phase more loops are produced at shorter lengths. This already suggests interesting implications for the calculation of the gravitational-wave spectrum (\cref{sec:gw}), which depends on the loop number density. The right panel of \cref{fig:n} features a gap which arises due to the fact that we assume an instantaneous
switch from the steady current attractor to the standard attractor
for the long string network.
This means that the loop size at formation is suddenly changed from
$\alpha_L \alpha(t_\text{last})$ to $\alpha_L \alpha$,
with $\alpha > \alpha(t_\text{last})$. In the realistic case of a current gradually turning on and off, the gap will be partially filled, thus increasing the gravitational-wave signal coming from the current--carrying phase. In this sense, our assumption of instantaneous turn off yields a conservative estimate of the spectrum.

 \subsection{Energy density of the network}
\label{sec:energy_density}

Let us now move on to discuss how the different components of the energy density related to the cosmic string
network behave compared to the critical density.
In the standard scaling regime ($\delta=0$), both the long strings and
the loops constitute a constant and subdominant component
of the energy density at all times.
This is, however, no longer true in the presence of currents.

For the long strings, we have
\begin{equation}\label{eq:longstrings}
	\frac{\rho_\infty}{\rho_c} \sim G t^2 \frac{\mu}{L^2}
	= \frac{G \mu}{\alpha(t)^2}
	\simeq \frac{G \mu}{\alpha_s^2}
	\left(\frac{t}{t_{\mathrm{ini}}}\right)^{-2\delta}\,,
\end{equation}
where we have taken the critical density as $\rho_c \sim 1/Gt^2$. This shows that, given negative $\delta$ for steady current attractor (see again \cref{sec:steady}), the energy density in the long strings has a power-law growth with the duration of the steady-current phase $t/t_{\mathrm{ini}}$.

The time evolution of the energy density in loops is given by
the following integral
\begin{equation}\label{eq:rhoLc}
	\frac{\rho_L}{\rho_c} \sim G t^2 \int
	\dd{\ell} \,\mu \, \ell \, n(\ell,t)\,,
\end{equation}
which can be evaluated assuming $\alpha_L\alpha(t)\gg \Gamma G \mu$ by using \eqref{eq:t4nfinal} for the radiation era.
The result becomes particularly
simple when we look at sufficiently late times,
$t\gg t_{\mathrm{ini}}/\Gamma G \mu$, for which we find\footnote{Notice that this is compatible with
$\alpha_L\alpha(t)\gg \Gamma G \mu$, which requires
$t \ll  t_{\mathrm{ini}}/(\Gamma G \mu)^{1/|\delta|}$.}
\begin{equation}\label{eq:ratiopp}
	\frac{\rho_L}{\rho_c} \sim
	\frac{\tilde c \mathcal{F} \kappa \alpha_L^2}{(1+\delta)(\alpha_L \alpha_s)^{\frac{3 \nu}{1+\delta}}}
	\frac{(G \mu)^{3+p} \Gamma^{2+p}}{(1+p)(2+p)} \left(\frac{t}{t_{\mathrm{ini}}}\right)^{-\frac{3 \nu \delta}{1+\delta}}\,.
\end{equation}
Here, $p=-4+3\nu/(1+\delta)\approx -2.2$.

When comparing the energy in loops with the energy in the long
strings we obtain
\begin{equation}\label{eq:Linf}
	\frac{\rho_\infty}{\rho_c} \left(\frac{\rho_L}{\rho_c}\right)^{-1}
	\sim (G \mu)^{-2-p} \left(\frac{t}{t_{\mathrm{ini}}}\right)^{-2\delta + 3 \nu \frac{\delta}{1+\delta}}
	\sim (G \mu)^{0.214\dots}
	\left(\frac{t}{t_{\mathrm{ini}}}\right)^{0.034\dots}\,.
\end{equation}
We thus conclude that given the very mild time growth in \eqref{eq:Linf} and the small prefactor, the energy density in loops will generically
be dominant with respect to the long string network.
For completeness, note that in the case with no currents
and $\delta=0$, we recover the standard results
$\rho_\infty/\rho_c \sim G \mu$ and $\rho_L/\rho_c \sim (G \mu/\Gamma)^{1/2}\,$.

Finally, let us evaluate the energy density in gravitational waves
during the current-carrying phase.
Each loop is expected to emit with a constant power
$P=\Gamma G\mu^2$ during its lifetime, so that the total amount of energy
in gravitational waves at the time $t$ is given by
\begin{equation}\label{eq:rhoR}
	\frac{\rho_\mathrm{GW}}{\rho_c} = \Gamma (G\mu)^2 t^2
	\int_{t_{\mathrm{ini}}}^t \left[
		\frac{a(t^\prime)}{a(t)}\right]^4 \mathcal{N}(t^\prime) \,\dd{t^\prime},
\end{equation}
where $\mathcal{N}(t)$
is the total number density obtained after integrating
$n(\ell,t)$ over all possible loop lengths,
$\mathcal{N}(t) = \int n(\ell,t) \,\dd{\ell}\,$.
The lower integration bound, $t_{\mathrm{ini}}$, ensures that
we select contributions only from loops formed after the occurrence
of the current.
 Loops formed prior to the current carrying
phase, as well as loops formed afterwards, can be accounted for
in the standard way.

\begin{figure}
	\centering
	\includegraphics[width=0.658\textwidth]{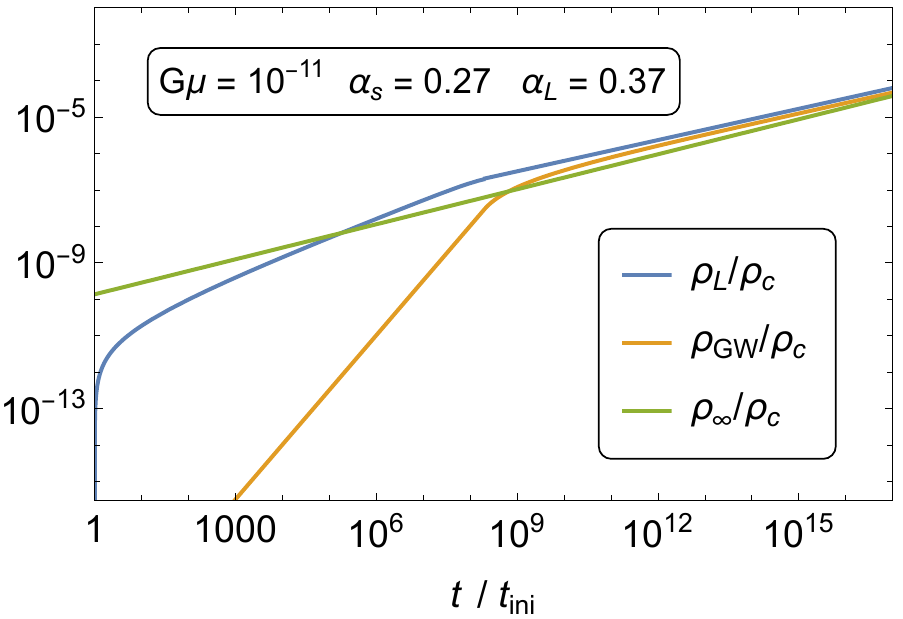}
	\caption{Energy density in small loops, $\rho_L$,
		gravitational waves, $\rho_\mathrm{GW}$, and long string network,
		$\rho_\infty$, normalized to the critical density $\rho_c \sim 1/Gt^2$, as a function of time after the onset of the current--carrying phase, $t_\text{ini}$.
		}
	\label{fig:rhos}
\end{figure}

Direct evaluation of \eqref{eq:rhoR} at times
$t\gg t_{\mathrm{ini}}/\Gamma G \mu$
shows that the energy density in loops and in gravitational waves
are proportional to each other and of the same order,
\begin{equation}
	\frac{\rho_\mathrm{GW}}{\rho_L}
	\simeq \frac{(1+\delta)(-2-p)}{-2-2\delta+4\nu+\delta\nu} \sim \mathcal{O}(1)\,.
\end{equation}
A summary plot with the behavior of the different components
of the energy density associated to the cosmic strings is shown in \cref{fig:rhos}.
We conclude that the most relevant contribution to the total
energy density comes from the population of small loops and it grows in time.
This can still be a subdominant component provided that
the right hand side of \eqref{eq:ratiopp} is much smaller than unity
at the time at which the current-carrying phase has ended,
providing agreement with standard cosmology.
The disappearance of the current can occur because of a change in the microscopic properties
of the system,
\eg the occurrence of a phase transition (see the comment on RHN currents in the Introduction), or due to the onset of matter domination in which the steady current attractor no longer exists.

Let us close this section by stating that, in the current-carrying scenario, the appearance of vortons is anticipated
\cite{Davis:1988ij, Brandenberger:1996zp, Martins:1998gb, Carter:1999an, Lemperiere:2003yt, Auclair:2020wse, Battye:2021sji}.
However, the originality of the steady-current attractor is the combination of a high current and of a reduced loop formation size, which favor the formation of vortons.

One may be concerned that the abundance of vortons $\rho_\mathrm{vort}$ could become dominant and overclose the Universe.
We address this concern by bounding $\rho_\mathrm{vort}$ from above with the extreme scenario where the loops produced by the network become vortons instantly, \emph{i.e.} $\Gamma = 0$ in \cref{eq:lnd}.
Performing an integral analogous to \cref{eq:rhoLc}, we find that the abundance of vortons is bounded by
\begin{equation}
    \frac{\rho_\mathrm{vort}}{\rho_c} \ll \frac{\tilde{c} \mathcal{F} \kappa G\mu}{\alpha_s^2 \gamma (2 + 2\delta - 3\nu)} \left(\frac{t}{t_\mathrm{ini}}\right)^{2-3\nu}.
\end{equation}
Therefore, in radiation era $\nu=1/2$, the steady-current attractor needs to last at least until $t \sim (G\mu)^{-2} t_\mathrm{ini}$ for the vortons to dominate the energy density of the Universe. However, as we assume a finite lifetime of the current propagating on strings, reaching this point can be easily avoided.

Another possible concern could be that vortons may reduce the amount of GWs produced. It has been shown quantitatively in Ref.~\cite{Auclair:2020wse}, in the context of standard scaling, that this does not happen. Indeed, the loops produced by the network during scaling need to lose nearly all of their energy to GWs before reaching a size small enough to become stable vortons. In the steady-current attractor solution, the situation is less clear due to the high currents on the string. In this work we assume that the conclusions of Ref.~\cite{Auclair:2020wse} still hold and hence that vortons are not relevant for the GW signature. If this were not the case, then the vortons would become unstable when the current leaks at $t_\mathrm{end}$ and resume GW production. We stress that this scenario deserves a more thorough and quantitative analysis in future work.

\section{Gravitational wave signal}
\label{sec:gw}

\noindent
Equipped with the modified loop number density during the current-carrying phase we are now ready to present the gravitational wave spectra emitted by CCCSs. We work with the assumption that the current-carrying phase does not match the time of string network formation. In particular, it is expected
that the current does not form at the time of the phase transition, but only at a later time when the temperature of the thermal bath has become sufficiently low. This is because the temperature needs to be smaller than the difference in current-carrier  mass inside and outside of the string.

We  describe the gravitational wave signal
as the superposition of three different contributions,
\begin{equation}
	\Omega_\mathrm{GW}^{tot}= \Omega_\mathrm{GW}^0(t_F,t_{\mathrm{ini}}) +
	\Omega_\mathrm{GW}(t_{\mathrm{ini}},t_\text{end})+
	\Omega_\mathrm{GW}^0(t_\text{end},t_0)\,,
\end{equation}
stemming from

\begin{enumerate}[label=($\roman*)$]
	\item loops whose formation time is $t_F<t<t_{\mathrm{ini}}$,
	      where $t_F$ is the formation time of the cosmic string
	      network (which can be approximately taken as the time
	      of the corresponding phase transition)
	       and $t_{\mathrm{ini}}$ is the
	      time at which the current first appears. For these loops
	      we adopt the standard loop production function\,;

	\item loops formed during the current-carrying phase,
	      for which the loop number density is given in \eqref{eq:t4n}.
	      This will be the case for $t_{\mathrm{ini}}<t<t_\text{end}$, where
	      $t_\text{end}$ corresponds to the time at which the current on
	      the network disappears either because of leakage effects or because
	      the Universe entered matter domination era (as discussed in
	       \cref{sec:steady-stability}, there is no steady current attractor solution
	       during such times)\,;

	\item loops formed after the end of the current carrying phase,
	      $t>t_\text{end}$, for which the production function is again
	      taken to be the standard one as in $(i)$. We argue that this is
	      a conservative approach because the system will not instantaneously enter
	      the scaling solution; instead, there will be a period of enhanced loop production
	      right after the current is switched off\,.
\end{enumerate}

\begin{figure}
	\centering
	\includegraphics{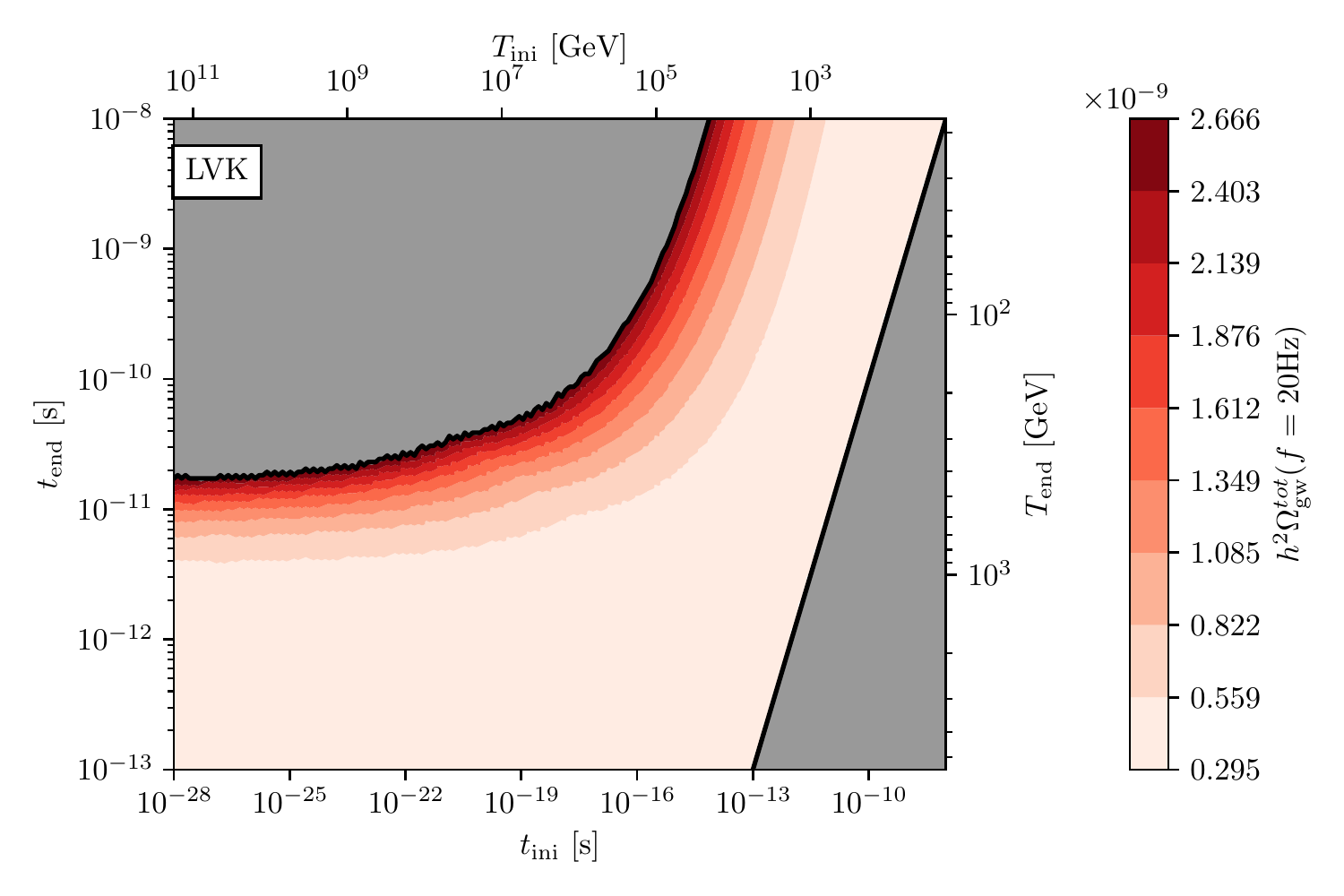}
	\caption{GW spectrum as a function of the time at which the current appears, $t_{\mathrm{ini}}$, and disappears, $t_\text{end}$, from the network
	for a string tension of $G \mu = 5\times 10^{-11}$. The parameter region labeled LVK is disfavored by searches for the stochastic gravitational wave background at ground-based interferometers \cite{KAGRA:2021kbb}.
    We excluded GW signals with $h^2 \Omega_\mathrm{GW} > 5\times 10^{-9} h^2$ at $20$\,Hz corresponding to the limits set on a flat power spectrum with a log-uniform prior.
    }
	\label{fig:time-time}
\end{figure}

For contributions from $(i)$ and $(iii)$ we use the standard formula for energy density in gravitational waves, $\Omega_\mathrm{GW}^0$, see for instance Eq.~(27) in Ref.~\cite{Gouttenoire:2019kij}. Hence, we will only show here the expression for gravitational wave energy density
during current-carrying phase. Note that the average power spectrum for GW emission reads
$P_k=\Gamma/(k^q \zeta(q))$, and we focus on cusps for which $q=4/3$. In this work we take $\Gamma=50$;
we stress that in a recent work \cite{Rybak:2022sbo}, the authors established a phenomenological relation between
$\Gamma$ and the value of the current. It turns out that by fixing $\Gamma$ our prediction for the modification of the gravitational wave spectrum in the current-carrying phase is on the conservative side.

Utilizing results from \cref{sec:loop_number}, we obtain the following formula for $k$-th mode contribution to the gravitational wave spectrum
\begin{equation}
\Omega_\mathrm{GW}^{k}(t_{\mathrm{ini}},t_\text{end}) = \frac{2k \Gamma^k G \mu^2}{\rho_c f} \frac{\tilde{c} \mathcal{F} \kappa}{\gamma} \int_{t_{\mathrm{ini}}}^{t_0} \frac{\theta(t_*^k-t_{\mathrm{ini}})\theta(t_\text{end}-t_*^k) (t_*^{k})^{-4} \dd{t}}{\alpha_L \alpha(t_*^{k})^3[\Gamma G\mu + \alpha_L \alpha(t_*^k)(1+\delta)]} \qty[\frac{a(t)}{a_0}]^5 \qty[\frac{a(t_*^k)}{a(t)}]^3
\label{eq:GWspectrum}
\end{equation}
where  $t_0$ denotes the present time and $t_*^k$ is the formation time of the loop contributing to the production of gravitational waves of frequency $f$ at time $t$.
Note that the Heaviside functions $\theta(t_*^k-t_{\mathrm{ini}})\theta(t_\text{end}-t_*^k)$ only require that the loops \emph{were formed} during the steady current attractor.
Therefore, some of these loops continue to emit gravitational waves after the end of the steady current attractor.
The formation time $t_*^k$ is the solution of

\begin{equation}\label{eq:ltt}
	\frac{2 k a(t)}{f a(t_0)} = \alpha_L \alpha(t_{\mathrm{ini}})(t_*^k/t_{\mathrm{ini}})^\delta t_*^k -
	\Gamma G \mu(t-t_*^k)\,,
\end{equation}
for which a good approximation is given by
\begin{equation}
	t_*^k \approx \left\{ \frac{t_{\mathrm{ini}}^\delta}{\alpha_L \alpha(t_{\mathrm{ini}})}
		\left[\frac{a(t)}{a(t_0)} \frac{2 k}{f} + \Gamma G \mu t\right]
		\right\}^{\frac{1}{1+\delta}}.
\end{equation}
In order to compute the gravitational wave spectra, we have also set $F=0.1$, $\gamma=1$, $\alpha_L=0.37$, $\alpha_s=0.27$, which are the values commonly used in the literature.

In passing, let us stress that, as expected, the general expression in \cref{eq:GWspectrum} matches the standard one in the limit $\delta \to 0$.
In order to compute the GW spectrum, we sum over $10^6$ modes, namely,
\begin{equation}
\Omega_\mathrm{GW}(t_{\mathrm{ini}},t_\text{end})=\sum_{k=1}^{10^6}\Omega_\mathrm{GW}^{k}(t_{\mathrm{ini}},t_\text{end})\,.
\end{equation}
We stress that the current is expected to cause a backreaction on string loops. This can lead to the emission of current carriers from strings cusps, which would result in the smoothing of the cusps. We, therefore, in addition to the full contribution from all cusps also compute the case where string loops oscillate at their fundamental frequency  (only taking mode $k=1$).
Hence, for each of the five benchmark cases in \cref{fig:money} we show both $k=1$ and $k\lesssim 10^6$ contributions and this is what gives a ``width'' to the predicted gravitational wave spectra.

What is immediately obvious by looking at each of the spectral predictions is
an enhancement in $\Omega_\mathrm{GW}^{tot}$ arising from the period when the current is on. This enhancement can be explained by looking at the denominator of \cref{eq:GWspectrum}. For $-1<\delta<0$,  ${\alpha_L \alpha_s [\alpha_L \alpha_s(1+\delta) (t_*^k/t_{\mathrm{ini}})^\delta + \Gamma G\mu]}$ would attain smaller values than in the standard $\delta=0$ case, leading to the rise in the spectrum. The origin of the enhancement is the increased loop number density in presence of a current when compared to the standard case.

In \cref{fig:money} we consider five different scenarios (denoted with (A)--(E)), corresponding to different values of $t_{\mathrm{ini}}$ and $t_\text{end}$ and this is what causes the rise of the spectrum in different frequency regions. Benchmark points (A), (C) and (E) are computed for $G\mu=0.5\times 10^{-10}$. For benchmark point (A), $t_\text{end}$ is the time of QCD phase transition; further, the current for point (C) ends at the time of electroweak phase transition while for (E) the current stops flowing at the time that is $10^{-4}$ times smaller than for (C). As the current gets turned off earlier, the spectral enhancement shifts to higher frequency. Benchmark points (A) and (C) are receiving current-induced  spectral modification in the frequency region where ground-based interferometers such as LIGO \cite{Shoemaker:2019bqt,LIGOScientific:2021nrg} are most sensitive. In addition, (A) can also be tested at forthcoming space-based interferometers such as BBO \cite{Corbin:2005ny} and DECIGO \cite{Kawamura:2020pcg}. Benchmark point (E), on the other hand, is interesting in the context of experiments sensitive to GWs in MHz region \cite{Aggarwal:2020olq}.
The two remaining benchmark points are calculated for $G\mu=0.5\times 10^{-17}$. For (B) and (D), the current stops at QCD and electroweak phase transition, respectively. As $G\mu$ gets smaller, we observe that for a fixed $t_\text{end}$ the spectral enhancement is shifted to higher frequencies; we can see that by comparing (A) and (B) as well as (C) and (D) for which current stops flowing at the same time.

We point out that the enhancement in the spectrum in the region where PTA experiments such as NANOGrav \cite{NANOGrav:2020bcs} are sensitive is not supported by  standard cosmology; the same holds for LISA \cite{LISA:2017pwj,LISACosmologyWorkingGroup:2022jok}. Namely, to make those experiments sensitive to the considered CCCS scenario, one would need to require that currents last for a very long time, which can easily result in an overclosure problem or unwanted modifications of the CMB power spectrum. In addition, it is not clear whether any microscopic theory of current generation and decay would realistically allow one to delay the time of current quenching to very late times. Naively, we would expect phase transitions in the string network to occur much earlier.
We nevertheless observe that a portion of the parameter space of cosmic string tension that is in the standard case considered to be unreachable by ground-based detectors such as LIGO, becomes testable in the scenario where cosmic strings feature currents. For all benchmark points in \cref{fig:money} we use
$t_{\mathrm{ini}}=10^{-35}$ s. In \cref{fig:time-time} we show the dependence of the  total gravitational wave spectrum as a function of $t_{\mathrm{ini}}$ and $t_\text{end}$
fixing the string tension as $G \mu = 10^{-11}$.
The region excluded by present ground-based interferometers
is indicated in the upper left corner as ``LVK''.
Notice that for $t_\text{ini} < 10^{-28}$ s the exclusion line becomes flat and the only non trivial dependence is on $t_\text{end}$, which controls the time at which the network is assumed to be back in scaling. This is because further lowering $t_\text{ini}$ will enhance the signal, but only at frequencies outside the LVK sensitivity.


\section{Conclusions}
\label{sec:conclusions}


The emission of gravitational waves by cosmic-string networks in the early Universe is typically studied in terms of simple models, \textit{e.g.}, the Nambu--Goto action or the Abelian Higgs model.
In this work, we made an attempt to go beyond the standard Nambu--Goto approximation by investigating the implications for the gravitational-wave spectrum in the presence of additional worldsheet degrees of freedom, namely, charge carriers that give rise to a current flowing on cosmic strings.
We specifically focused on neutral currents, \ie, currents composed of particles that do not interact with any long-range force field, in order to avoid energy loss of the string network via other decay channels on top of the emission of gravitational waves.
At the same time, we restricted ourselves to the case of chiral currents, which are characterized by a particularly simple state parameter $K = Q^2 - J^2 = 0$ and which appear to be well motivated in view of the possibility of chiral RHN currents on cosmic strings in $SO(10)$ GUT models.


In order to study the evolution of a current-carrying cosmic string (CCCS) network, we employed the generalized velocity-dependent one-scale model formulated in  Ref.~\cite{Martins:2020jbq}, which allows one to track the time dependence of three characteristic properties of the network: its correlation length, the RMS velocity of long strings, and the strength of the current carried by the network.
After rewriting them as a nonlinear autonomous system, we analyzed and solved the generalized VOS equations, which allowed us to identify two attractor solutions:
one solution in the vicinity of the standard VOS attractor solution featuring a decaying current, $Y \rightarrow 0$;
as well as a new solution in the large-current regime, $Y \rightarrow 1$, featuring a decaying dimensionless correlation length $\alpha = L/t$ as well as a decaying RMS velocity $v$.
The second attractor solution can be reached whenever the process of current formation on cosmic strings results in a large initial current that is of $\mathcal{O}\left(1\right)$ right from the start; see \eg the $Y_{\rm min}$ values in Tab.~\ref{table:steady}.
In future work, it will therefore be interesting to identify concrete microscopic CCCS models that can give rise to such large currents on cosmic strings, \ie, physical current energy densities as large as $\mathcal{O}\left(\mu\right)$.


For the purposes of this paper, we decided to restrict ourselves to a bottom-up phenomenological analysis and focus on the implications of the second attractor solution of the generalized VOS model for the gravitational-wave spectrum.
To this end, we computed the loop number density as well as energy density of a CCCS network based on the $Y\rightarrow 1$ attractor solution, which led us to several interesting observations.
First of all, the shrinking dimensioneless correlation length $\alpha = L/t$ during the current-carrying phase results in the production of smaller loops, which boosts the loop number density in the small-loop regime and causes a gap in the loop number density after the end of the current-carrying phase; see Fig.~\ref{fig:n}.
At the same, the steadily decreasing value of $\alpha$ results in a growing fractional energy density of long strings, $\Omega_{\infty} = \rho_\infty / \rho_c \propto \alpha^{-2}$, which eventually translates into growing fractional energy densities of loops and gravitational waves; see Fig.~\ref{fig:rhos}.
The extra energy in gravitational waves notably manifests itself in a peak-like enhancement of the standard gravitational-wave spectrum from cosmic strings whose boundaries are controlled by the times of current generation and decay; see Fig.~\ref{fig:money}.


The boosted gravitational-wave signal across large ranges in frequency implies interesting prospects for current and upcoming gravitational-wave experiments.
Values of the cosmic-string tension that are out of reach of existing experiments may \eg be probed in the near future if currents should be responsible for an enhanced signal in the right frequency band; see \eg benchmark points (A) and (C).
At the same, we observe that gravitational waves from CCCSs are an interesting target for planned gravitational-wave searches at high frequencies, reaching up to the MHz regime; see \eg benchmark point (E).
Finally, we caution that our results come with a sizable uncertainty, which is, at least to some extent, reflected in the width of the different bands in Fig.~\ref{fig:money}.
Ultimately, our results call for an independent confirmation based on numerical simulations.
The validity of the standard VOS model has been well confirmed by simulations; similarly, simulations of CCCS networks might allow one to confirm the validity of the generalized VOS model, and hence our promising predictions for the associated gravitational-wave signal.


\section*{Acknowledgements}


We thank Patrick Peter, Christophe Ringeval, Paul Shellard and  Lara Sousa for useful discussions. Fermilab is managed by Fermi Research Alliance, LLC (FRA), acting under Contract No.\ DE-AC02-07CH11359.
The work of PA is partially supported by the Wallonia-Brussels Federation Grant ARC \textnumero~19/24-103.
The work of SB is supported by the Strategic Research Program High-Energy Physics and the Research Council of the Vrije Universiteit Brussel, and by the
``Excellence of Science - EOS" - be.h project n.30820817.


\bibliographystyle{JHEP}
\bibliography{refs.bib}


\end{document}